\documentclass[epj,nopacs]{svjour}
\usepackage{graphicx}
\usepackage{amsmath}
\usepackage{amsfonts}

\newcommand{\beq}{\begin{equation}}
\newcommand{\eeq}{\end{equation}}
\newcommand{\beqa}{\begin{eqnarray}}
\newcommand{\eeqa}{\end{eqnarray}}
\def\lsim{\raise0.3ex\hbox{$<$\kern-0.75em\raise-1.1ex\hbox{$\sim$}}}
\def\gsim{\raise0.3ex\hbox{$>$\kern-0.75em\raise-1.1ex\hbox{$\sim$}}}

\def\0{{\bf 0}}

\begin{document}

\title{Heavy flavours in heavy-ion collisions: quenching, flow and correlations}

\author{A. Beraudo \and 
A. De Pace \and M. Monteno \and 
M. Nardi \and F. Prino}

\institute{Istituto Nazionale di Fisica Nucleare, Sezione di Torino, \\ 
  via P.Giuria 1, I-10125 Torino, Italy}

\date{}

\abstract{We present results for the quenching, elliptic flow and azimuthal correlations of heavy flavour particles in high-energy nucleus-nucleus collisions obtained through the POWLANG transport setup, developed in the past to study the propagation of heavy quarks in the Quark-Gluon Plasma and here extended to include a modeling of their hadronization in the presence of a medium. Hadronization is described as occurring via the fragmentation of strings with endpoints given by the heavy (anti-)quark $Q(\overline{Q})$ and a thermal parton $\overline{q}(q)$ from the medium. The flow of the light quarks is shown to affect significantly the $R_{AA}$ and $v_2$ of the final $D$ mesons, leading to a better agreement with the experimental data. The approach allows also predictions for the angular correlation between heavy-flavour hadrons (or their decay electrons) and the charged particles produced in the fragmentation of the heavy-quark strings.}

\maketitle

\section{Introduction}
Heavy quarks -- indirectly accessible through $D$-mesons, heavy-flavour decay electrons and muons and $J/\psi$'s from $B$ decays -- have been used for a long time as probes of the medium formed in heavy-ion collisions. Produced in hard processes during the crossing of the two nuclei, they traverse the fireball until decoupling from the latter. During their path in the plasma they lose energy, they tend to inherit part of the flow (radial and elliptic, in non-central collisions) of the medium and to decorrelate from their companion coming from the same hard interaction. The first experimental results in Au-Au collisions at $\sqrt{s_{\rm NN}}\!=\!200$ GeV at RHIC, although limited to the non-photonic electrons from semileptonic decays of $D$ and $B$ mesons measured by the PHENIX and STAR collaborations~\cite{PHEe,STARe}, were surprisingly displaying a sizable suppression (comparable to the one of pions) of the yields at high $p_T$ and a non negligible elliptic flow: this led people to wonder about the possibility of (kinetic) thermalization of charm quarks in the plasma, eventually flowing with the rest of the medium. 
If these first results could not be considered conclusive, because of the large uncertainty due to the background subtraction and the impossibility of disentangling the separate charm and beauty contribution, measurements in Pb-Pb collisions at $\sqrt{s_{\rm NN}}\!=\!2.76$ TeV at the LHC performed by the ALICE and CMS collaborations confirmed the above findings looking at a much wider set of observables: besides the electrons~\cite{ALICEe} and muons~\cite{ALICEmu} from heavy flavour hadron decays it was possible to measure over a wide $p_T$-range the $R_{AA}$~\cite{ALICE_DRAA} and $v_2$~\cite{ALICE_Dv2} of $D$ mesons and (for high enough $p_T$) the quenching of non-prompt $J/\psi$'s~\cite{CMS_Jpsi}. 
Recently the STAR collaboration at RHIC managed to measure the $R_{AA}$ of $D$ mesons down to very low $p_T$~\cite{STARD} with a very fine binning which, besides the quenching at high $p_T$, turned out to be characterized by a sharp enhancement around $p_T\!\sim\!1.5$ GeV/c whose origin may be attributed to the radial flow acquired by charm quarks in the plasma (in a scenario in which the $c$-quarks approach thermalization), to the recombination with light partons from the medium at the end of the QGP phase or to a combination of the two effects. 
A quantitative answer requires then to include in the theoretical calculations, besides a description of the propagation of the heavy quarks in the plasma, also a modeling of their hadronization in the presence of a medium.
Furthermore, measurements of azimuthal correlations ($e\!-\!h$, $D\!-\!h$, $e^+\!-\!e^-$...) in A-A collisions, sensitive to the in-medium decorrelation of $Q\overline{Q}$ pairs produced in the same hard interaction, might also become accessible in the next years and theory models should be capable of providing predictions for these observables.   

Various theoretical models were proposed in the past to describe the above findings, based on the Boltzmann, Fokker-Planck or Langevin equations~\cite{aic,aic2,BAMPS,BAMPS2,tea,aka,rapp,rapp2}.
In a series of papers~\cite{lange0,lange,lange2} over the last few years we developed a complete setup (referred to as POWLANG) for the study of heavy flavour observables in high-energy nucleus-nucleus collisions, describing the initial hard production of the $Q\overline{Q}$ pairs and the corresponding parton-shower stage through the POWHEG-BOX package~\cite{POW,POWBOX} and addressing the successive evolution in the plasma through the relativistic Langevin equation.
 Here we supplement our numerical tool by modeling the hadronization of the heavy quarks accounting for the presence of a surrounding medium made of light thermal partons feeling the collective flow of the local fluid cell. As we will show in this paper, this will have important effects on the heavy flavour hadron momentum distributions in the low $p_T$ region, since the collective flow of the light partons from the medium will be inherited by the final charmed hadrons: the $D$-meson $R_{AA}$ will then display an enhancement at moderate $p_T$ and also their elliptic flow will result increased, improving the agreement with the experimental data.

Our paper is organized as follows.
Sec.~\ref{sec:pp} is focused on p-p collisions, where we display the POWHEG-BOX predictions for the azimuthal $D$-hadron (and electron-hadron) correlations. The satisfactory agreement with the preliminary $D\!-\!h$ data measured by the ALICE collaboration (besides the one with the inclusive single-particle spectra) represents a further validation of the pQCD Monte Carlo tool employed to describe the initial $Q\overline{Q}$ production.
In Sec.~\ref{sec:AA} we move to the case of A-A collisions. In order to provide a theoretical benchmark we first illustrate in Sec.~\ref{sec:thermal} what would be the predictions of hydrodynamics in the extreme limit in which heavy-flavour particles reached full kinetic equilibrium with the medium. In Sec.~\ref{sec:model} we describe our model of in-medium heavy quark hadronization and its numerical implementation taking advantage of the PYTHIA string-fragmentation routine~\cite{PYTHIA}. In Sec.~\ref{sec:results} we display the results obtained interfacing the above hadronization routine to the outcomes of our POWLANG transport setup at the end of the heavy quark propagation through the medium; 
in particular we focus on the $R_{AA}$ and $v_2$ of $D$ mesons, both at RHIC and at the LHC. Heavy flavour azimuthal correlations in A-A collisions are addressed in Sec.~\ref{sec:corr}. Finally in Sec.~\ref{sec:discussion} we discuss the possible relevance of our findings for the interpretation of present and future experimental measurements.

\section{Heavy Flavour in p-p collisions: azimuthal correlations}\label{sec:pp}

The study of azimuthal correlations of $c\!-\!\bar{c}$ ($b\!-\!\bar{b}$) pairs is a useful tool to investigate the properties of the medium formed in high-energy nuclear collisions and its effects on the heavy-quark propagation (energy loss, angular decorrelation and possible modification of their fragmentation). It is accessible through the study of correlations of open-charm (beauty) hadrons and/or their decay products.
If results on direct $D\!-\!\overline{D}$ correlations in heavy-ion collisions cannot be realistically expected due to the small branching ratio into the hadronic decay channels used for their reconstruction, indirect information on the primordial $Q\overline{Q}$ pairs can be obtained correlating charged hadrons to a selected trigger particle from the same event, like a D-meson or an electron from heavy-flavour hadron decays ($D\!-\!h$ and $e\!-\!h$ correlations): this in particular, with proper kinematic cuts, should allow one to study possible medium modifications of the jet from the fragmentation of the correlated heavy quarks.  
Preliminary results on correlations between $D$ mesons and light hadrons have been presented by the ALICE Collaboration~\cite{sandro,rossi} (so far limited to p-p and p-Pb collisions) and analogous analyses using heavy-flavour decay electrons trigger particles have been also performed and are currently in progress~\cite{deepa}.

The above experimental information obtained in elementary p-p collisions can be exploited to validate the theoretical calculations used to simulate the initial hard $Q\overline{Q}$ production.  
In our setup, the heavy quarks are created in pairs by the \mbox{POWHEG-BOX} event generator. Their momenta are not back-to-back neither along the beam-axis, due to the different Bjorken-$x$ carried by the partons taking part in the hard event, nor in the azimuthal plane, due to the gluon radiation occurring during the hard process or the shower stage and also to the intrinsic $k_T$-broadening included in the simulation.
Eventually, heavy quark hadronization and the final decays of the $D$ ($B$) mesons are simulated with PYTHIA, which is also used to describe the parton shower stage.

\begin{figure}[!h]
\begin{center}
\includegraphics[clip,width=0.48\textwidth]{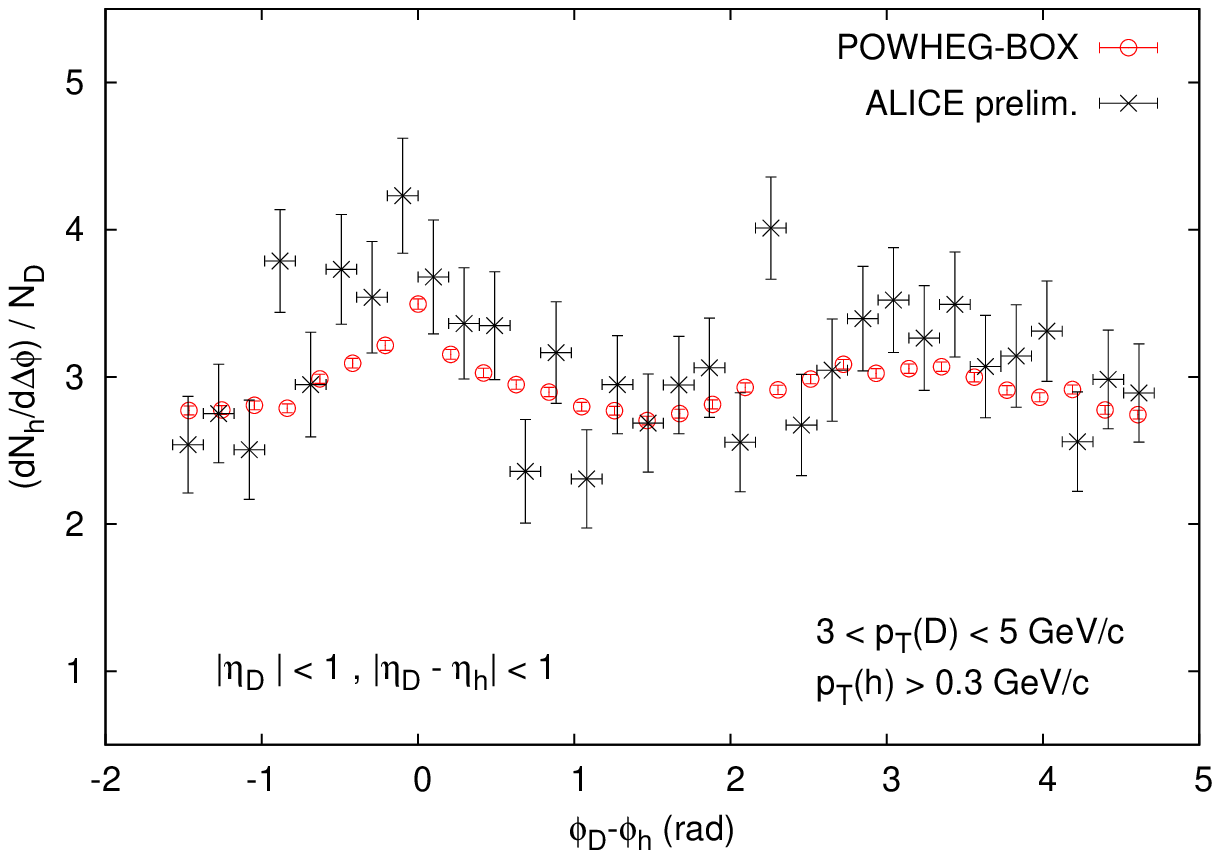}
\includegraphics[clip,width=0.48\textwidth]{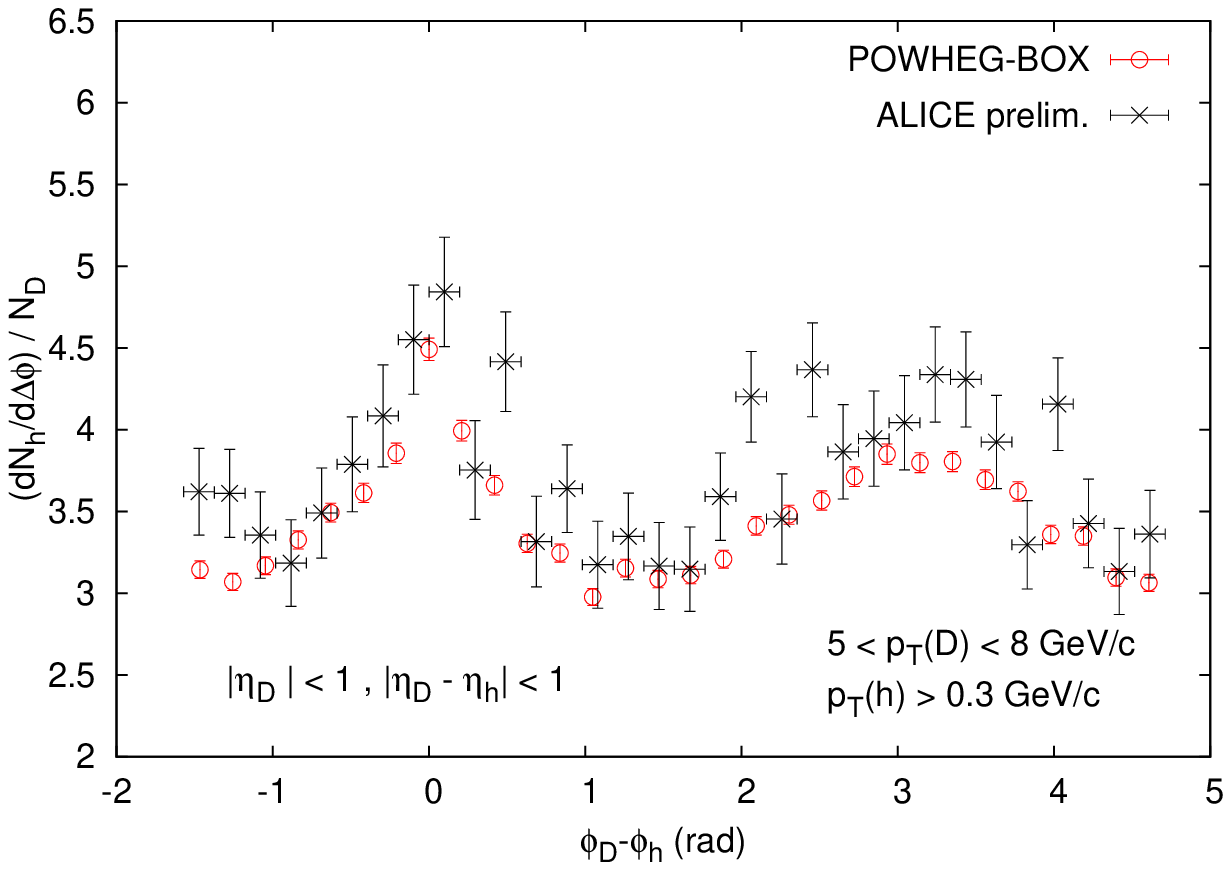}
\includegraphics[clip,width=0.48\textwidth]{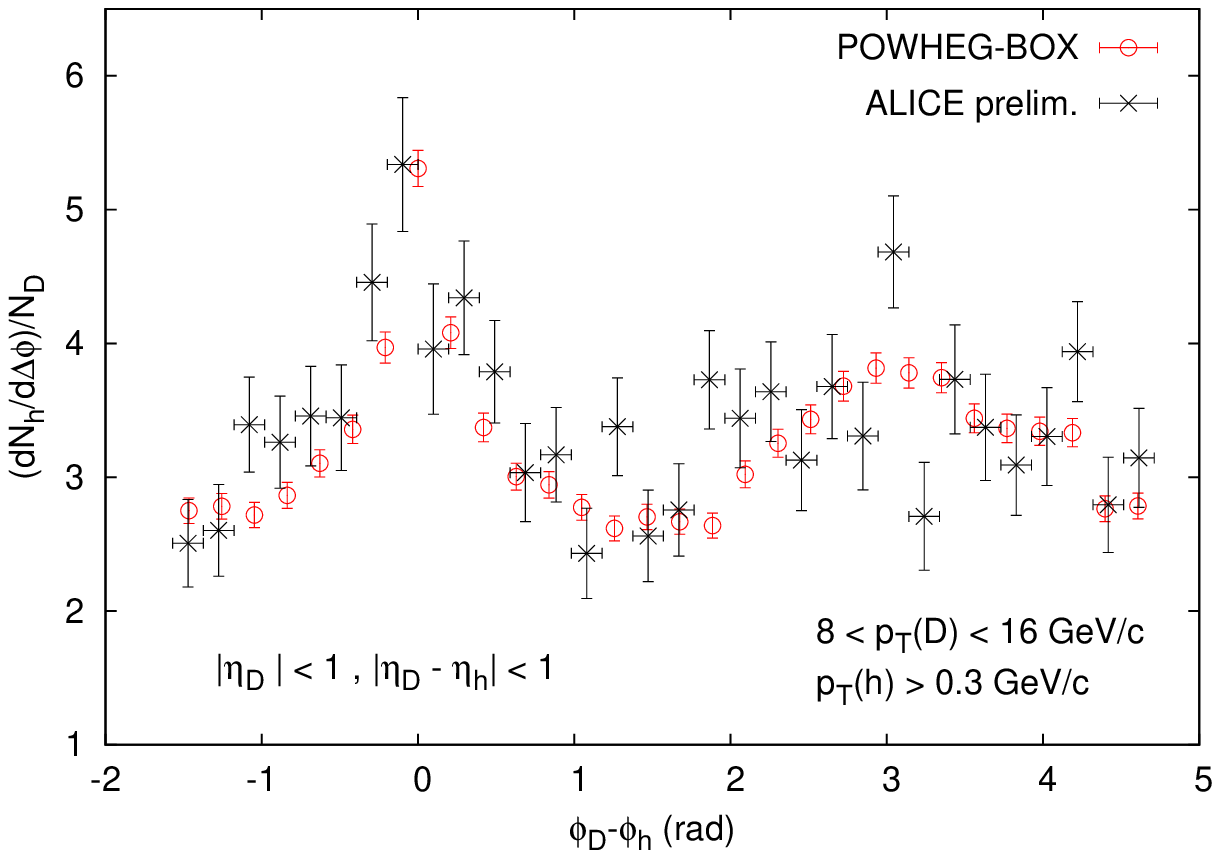}
\caption{Azimuthal $D\!-\!h$ correlations in pp collisions at $\sqrt{s}\!=\!7$ TeV for various cuts compared to preliminary ALICE data~\cite{sandro}.}\label{fig:ppcorr} 
\end{center}
\end{figure}

In Fig.~\ref{fig:ppcorr} we show our results for $D\!-\!h$ azimuthal correlations compared to preliminary ALICE data~\cite{sandro,rossi}, for three different $p_T$-intervals of the charmed meson. In our simulation $D^0$, $\bar{D}^0$ and $D^\pm$ are used as trigger particles and the light hadrons are limited to charged pions and kaons, protons and antiprotons, excluding the weak decays of $\Lambda$ and $K^0$. Any $D$ meson is correlated with all the light hadrons (except its own decay products) created in the same event.
The near side peak takes contribution both from correlations present at the partonic level (from $Q\overline{Q}$ pairs arising from gluon splitting) and from hadrons coming from the fragmentation of the same string of the parent heavy quark. Our results include also the simulation of the Underlying Event (UE) due to Multiple Parton Interactions (MPI), performed with PYTHIA 6.4, which gives rise to the pedestal observed in Fig.~\ref{fig:ppcorr}.

\begin{figure}[!h]
\begin{center}
\includegraphics[clip,width=0.48\textwidth]{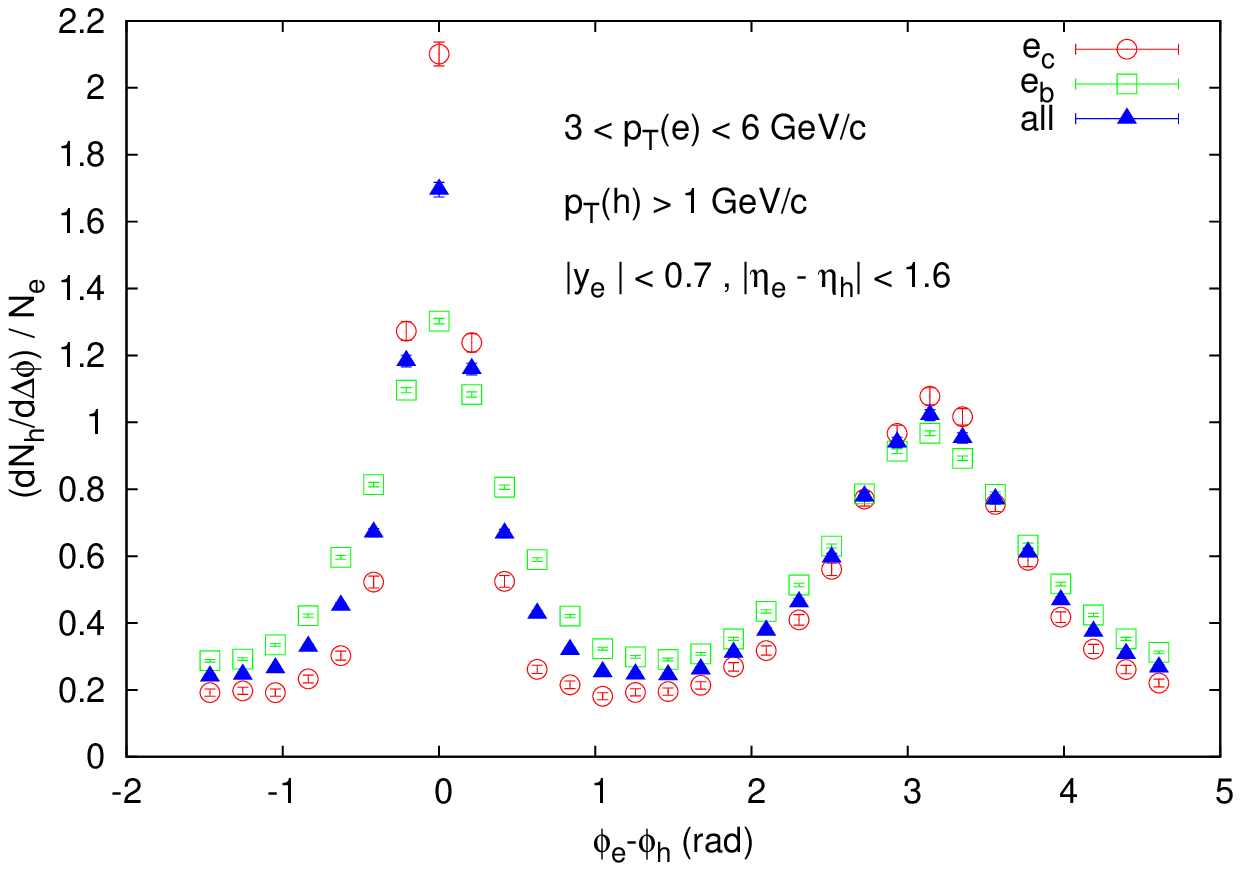}
\includegraphics[clip,width=0.48\textwidth]{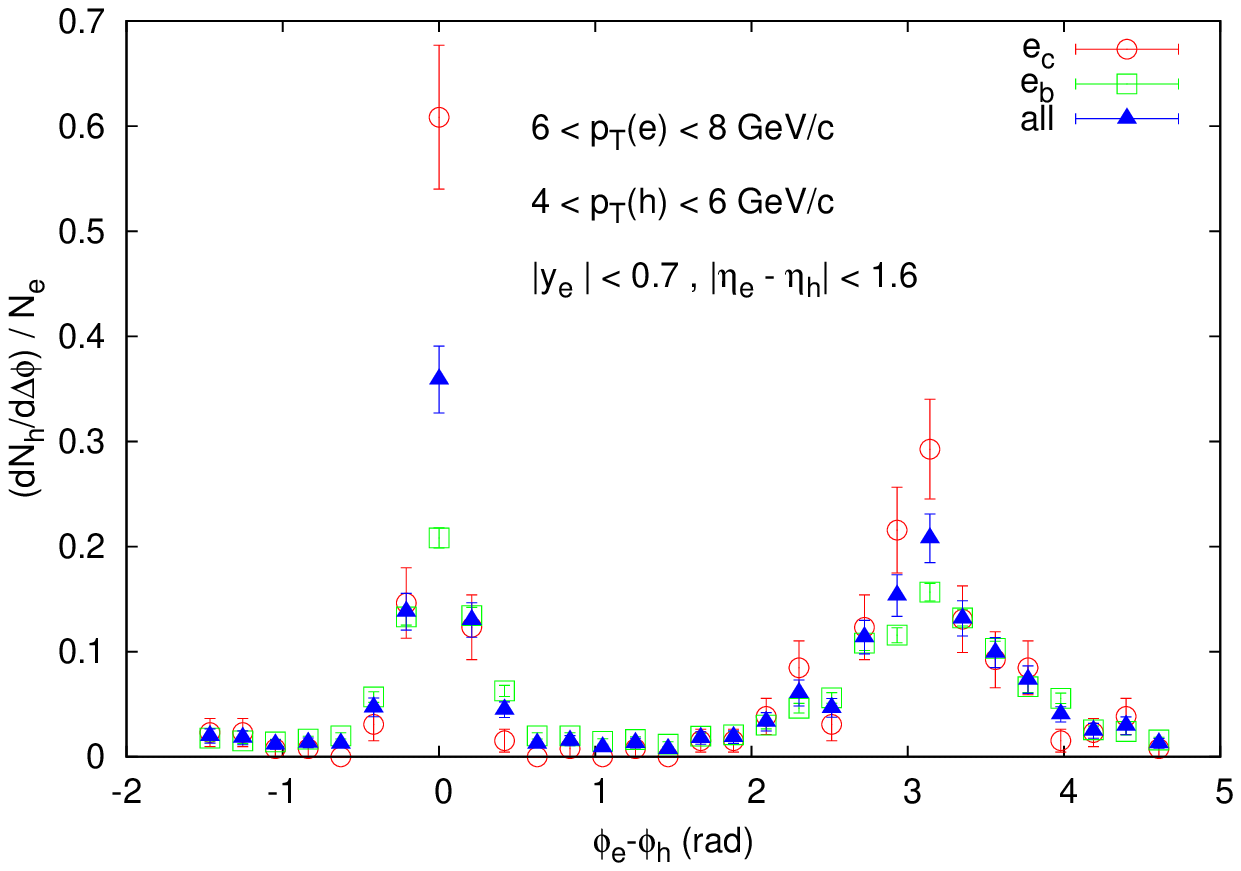}
\includegraphics[clip,width=0.48\textwidth]{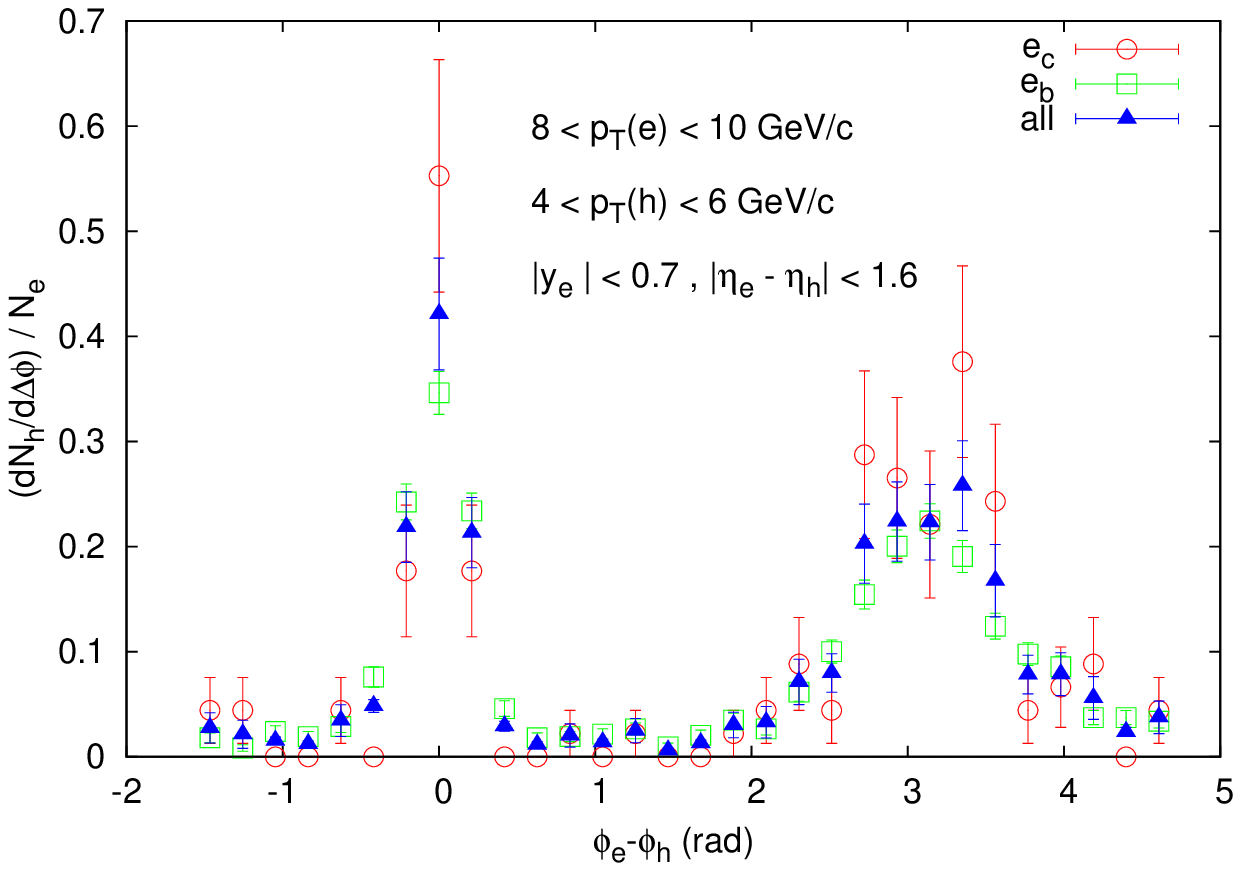}
\caption{Azimuthal $e\!-\!h$ correlations in p-p collisions at $\sqrt{s}\!=\!7$ TeV for various kinematical cuts accessible by ALICE.}\label{fig:ppe-hcorr} 
\end{center}
\end{figure}

A second channel to get indirect information on $Q\overline{Q}$ correlations is through the electrons from heavy-flavour hadron decays. Electron-hadron angular correlations are shown in Fig. \ref{fig:ppe-hcorr}. In this case, any trigger electron of heavy flavour origin is correlated with all light hadrons (limited as in the previuous case, for simplicity, to pions, kaons, protons and antiprotons) of the same events, including those coming from the parent $D/B$ mesons, which are not experimentally subtracted.
Furthermore, since experimental data include electrons coming both from $D$ and $B$ mesons, we have to consider both sources.
In our plots, red points show the correlations for electrons from $D$ decays, green points from $B$ decays (including those coming from the $B\to D \to e$ chain), and blue points refer to the average between them, weighted with the respective $Q\overline{Q}$ production cross-sections: $\sigma_{c\bar{c}}=4.131$ mb, $\sigma_{b\bar{b}}=0.237$ mb (as given by POWHEG) for p-p collisions at $\sqrt{s}=7$ TeV. To speed-up the calculation in this case no simulation of the UE was included: the small pedestal observed in Fig.~\ref{fig:ppe-hcorr} arises from soft hadrons decorrelated from their ancestors produced in the hard event. With all the adopted kinematic cuts the away-side peak is always clearly visible in p-p collisions.


\section{Heavy Flavour in A-A collisions}\label{sec:AA}
Having verified that we have at our disposal a validated tool to simulate the heavy-flavour hard production in p-p events, capable of providing a complete differential information on the final state, we now move to the A-A case.  
As done in Refs.~\cite{lange,lange2}, heavy quark propagation in the fireball formed in high-energy nuclear collisions will be simulated through a relativistic Langevin equation, solved in the presence of a medium described by viscous relativistic hydrodynamics~\cite{rom1,rom2}. Results will be presented for different choices of the heavy flavour transport coefficients in the Quark-Gluon Plasma: either from weak-coupling calculations with Hard-Thermal-Loop (HTL) resummation of medium effects or exctracted from lattice-QCD (lQCD) simulations~\cite{lat1,lat2,lat3}. At variance with Refs.~\cite{lange,lange2} different mechanisms will be tested to describe the transition from quarks to hadrons, either via standard vacuum fragmentation functions (FF) or through a routine based on string formation with light partons from the medium. The latter will be illustrated in Sec.~\ref{sec:model} (for a brief summary see also~\cite{ber}) and applied to the evaluation of $D$-meson $R_{AA}$ (also in-plane and out-of-plane) and $v_2$ in Sec.~\ref{sec:results}. Finally in Sec.~\ref{sec:corr} results for heavy-flavour correlations obtained with our new improved setup will be presented for the first time. However, before addressing full transport simulations, it is instructive to check what one would get in the extreme scenario in which charmed particles reached kinetic equilibrium with the rest of the fireball: this will be the subject of the next section.
\subsection{The thermal equilibrium scenario: Cooper-Frye spectra}\label{sec:thermal}
Since over the last few years experimental data on the heavy flavour $R_{AA}$ and $v_2$ have suggested the possibility that heavy quarks in nucleus-nucleus collisions may (at least partially) participate in the collective flow of the fireball, we will start considering the limiting scenario in which charmed particles reach full kinetic equilibrium with the medium. In this case the $D$ meson spectrum would be given by the same Cooper-Frye prescription~\cite{cooper} used for soft hadrons
\beq
E\frac{dN_D}{d^3p}=\frac{1}{(2\pi)^3}\int_{\Sigma_{\rm dec}}p\!\cdot\! d\Sigma_{\rm dec}\,\frac{1}{e^{p\cdot u(x)/T_{\rm dec}}-1},\label{eq:Cooper-Frye}
\eeq
where the decoupling temperature $T_{\rm dec}$ defines the isothermal freeze-out hypersurface for $D$ mesons. In the following the Cooper-Frye prescription is applied to a medium described by the hydrodynamic code by Luzum and Romatschke~\cite{rom1,rom2}, with initialization parameters typical of Au-Au and Pb-Pb collisions at RHIC and LHC: $\eta/s\!=\!0.08$, initial time $\tau_0\!=\!1$ and 0.6 fm/c and maximum temperature at the center $T_0\!=\!333$ and 475 MeV, respectively. For $T_{\rm dec}$ we will explore values around the QCD critical temperature, since we don't expect interactions in the hadronic phase (an interesting topic to explore in future work) to be able to maintain kinetic equilibrium: the spectrum provided by Eq.~(\ref{eq:Cooper-Frye}) would then reflect a thermalization of charm achieved in the partonic phase.  

\begin{figure}[!h]
\begin{center}
\includegraphics[clip,width=0.48\textwidth]{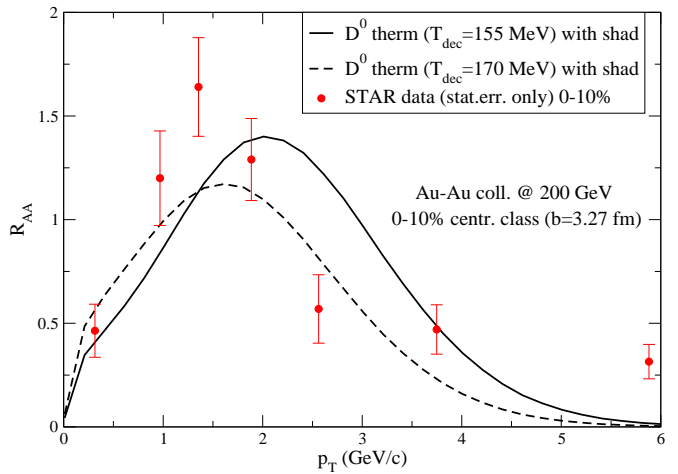}
\caption{$D^0$ $R_{AA}$ in central Au-Au collisions at $\sqrt{s_{NN}}\!=\!200$ GeV. Results for thermal $D^0$ spectra for two decoupling temperatures from the hydrodynamic evolution of the fireball (p-p and A-A spectra were normalized to give the same $Q\overline{Q}$ multiplicity, just rescaled by the effect of the nPDFs) are compared to STAR data~\cite{STARD}.}\label{fig:RAA_D_RHIC_therm} 
\end{center}
\end{figure}
We start such an investigation addressing the nuclear modification factor of $D^0$ mesons in Au-Au collisions at RHIC. For this observable the STAR experiment managed to provide data down to very low $p_T$~\cite{STARD}, displaying a characteristic enhancement (visible thanks to the very fine binning) around $p_T\sim 1.5$ GeV and challenging theoretical calculations to reproduce such a peculiar pattern. At first we test the extreme scenario of complete kinetic equilibration of charmed particles, which in this case should display the same radial flow of the medium. 
In Fig.~\ref{fig:RAA_D_RHIC_therm} we check whether this scenario allows one to describe the features observed in the $D^0$ $R_{\rm AA}$: STAR data obtained in Au-Au collisions at $\sqrt{s_{NN}}\!=\!200$ GeV in the $0-10\%$ centrality class are compared to the ``thermal'' $R_{\rm AA}$ arising from a Cooper-Frye decoupling at the temperatures $T_{\rm dec}=155$ and 170 MeV. 
Being the background medium described by a (2+1)D hydrodynamic flow (with longitudinal boost invariance) pp (provided by POWHEG-BOX) and AA spectra (per binary nucleon-nucleon collision) are normalized to give the same rapidity density around $y=0$, just rescaled by the change of the total $Q\overline{Q}$ cross section due to the effect of the nPDFs in the AA case. 
As it can be seen, the trend of the data looks in qualitative agreement with the hypothesis of kinetic equilibrium of charm, although a comparison with other observables should be performed and other mechanisms could be at work to give the observed modification of the $p_T$ spectrum: these items will be addressed in the following.  

\begin{figure}[!h]
\begin{center}
\includegraphics[clip,width=0.48\textwidth]{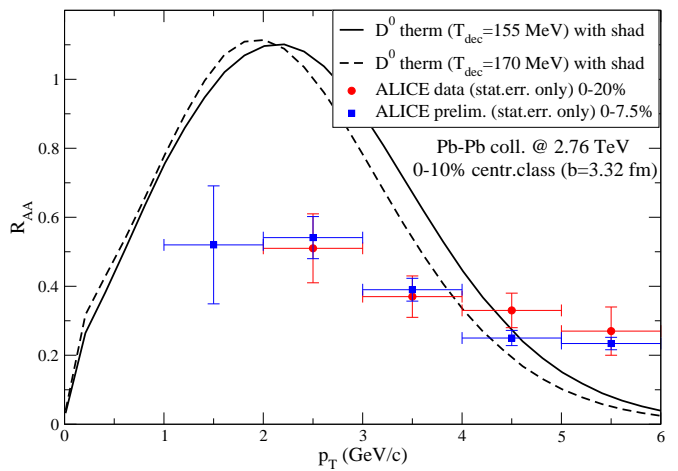}
\caption{The $D^0$ $R_{AA}$ in central Pb-Pb collisions at $\sqrt{s_{NN}}\!=\!2.76$ TeV. Results corresponding to kinetic thermalization of D mesons (effects of nPDFs are included in the normalization of the spectra) are compared to ALICE data for two different sets of central events~\cite{ALICE_DRAA,zaida}.}\label{fig:RAA_D_LHC_therm} 
\end{center}
\end{figure}
\begin{figure}[!h]
\begin{center}
\includegraphics[clip,width=0.48\textwidth]{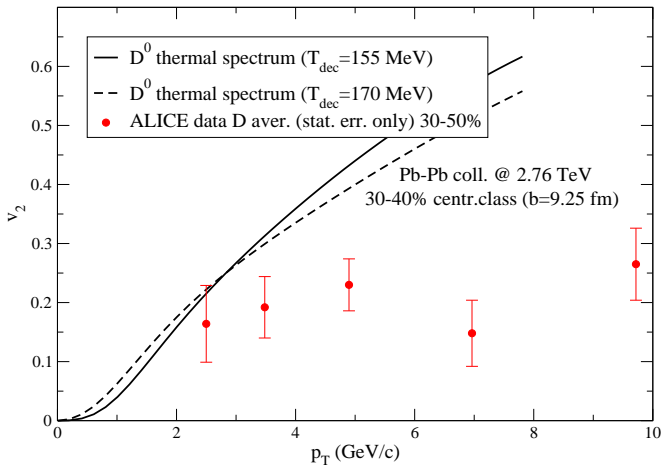}
\caption{The elliptic flow of $D$ mesons in semi-central ($b\!=\!9.25$ fm employed in the hydrodynamic evolution) Pb-Pb collisions at $\sqrt{s_{NN}}\!=\!2.76$ TeV. Results corresponding to kinetic thermalization of D mesons are compared to ALICE data~\cite{ALICE_Dv2} in the 30-50\% centrality class.}\label{fig:v2_D_LHC_therm} 
\end{center}
\end{figure}

In Figs.~\ref{fig:RAA_D_LHC_therm} and \ref{fig:v2_D_LHC_therm} we check whether the above scenario is able to describe also the data at the higher center of mass energy $\sqrt{s_{NN}}\!=\!2.76$ TeV reached in Pb-Pb collisions at the LHC. ALICE data for the $R_{AA}$~\cite{ALICE_DRAA} and $v_2$~\cite{ALICE_Dv2} of D mesons are compared to the corresponding theoretical predictions for the kinetic equilibrium case. As it can be seen the agreement with the theory curves in this case is less satisfactory, in particular for what concerns the $R_{AA}$, 
although the absence of experimental data in the more interesting region of low $p_T$ and the broader $p_T$ binning with respect to the one employed by the STAR Collaboration prevent us from drawing definite conclusions concerning this issue: notice in fact that so far (published) experimental data are limited to the region $p_T\gsim 2$ GeV (although preliminary results start to be available also for lower momenta), which also for light hadrons is at the border of the validity of hydrodynamic calculations. 
In Figs.~\ref{fig:RAA_D_LHC_therm} and \ref{fig:v2_D_LHC_therm} hydrodynamic predictions are given for impact parameters corresponding to the $0-10\%$ and $30-40\%$ more central events, while experimental data refer to a slightly different choice of centrality classes, which however is not expected to affect the above qualitative considerations.

\subsection{In-medium hadronization: a simple model}\label{sec:model}
\begin{figure}
\begin{center}
\includegraphics[clip,width=0.4\textwidth]{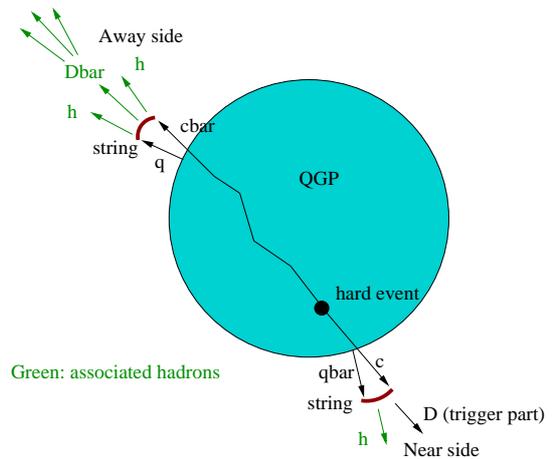}
\caption{A cartoon of the hadronization model interfaced to the POWLANG transport code at the end of the propagation of the heavy quarks in the plasma. We also illustrate how the information on the final particles from the string fragmentation can be used to study $D\!-\!h$ correlations (color on-line).}\label{fig:stringmodel} 
\end{center}
\end{figure}
In order to simulate the hadronization of heavy quarks in the medium at the end of their propagation in the QGP we proceed as follows. Once a heavy quark $Q$, during its stochastic propagation in the fireball, has reached a fluid cell below the decoupling temperature $T_{\rm dec}$, it is forced to hadronize. One extracts then a light antiquark $\overline{q}_{\rm light}$ (up, down or strange, with relative thermal abundancies dictated by the ratio $m/T_{\rm dec}$) from a thermal momentum distribution corresponding to the temperature $T_{\rm dec}$ in the Local Rest Frame (LRF) of the fluid; information on the local fluid four-velocity $u^\mu_{\rm fluid}$ provided by hydrodynamics allows one to boost the momentum of $\overline{q}_{\rm light}$ from the LRF to the laboratory frame. 
A string is then constructed joining the endpoints given by $Q$ and $\overline{q}_{\rm light}$ and is then passed to PYTHIA 6.4~\cite{PYTHIA} to simulate its fragmentation into hadrons (and their final decays). This is done as follows: the particle type, energy, polar and azimuthal angle of each endpoint are provided to PYTHIA through the PY1ENT subroutine; the PYJOIN subroutine allows one to construct the corresponding string; finally a PYEXEC call starts the simulation of its fragmentation and the final decays of unstable particles. In case the invariant mass of the string is not large enough to allow its decay into at least a pair of hadrons the event is resampled, extracting a new thermal parton to associate to the heavy quark. 
In agreement with PYTHIA, in evaluating their momentum distribution, light quarks are taken as ``dressed'' particles with the effective masses $m_{u/d}\!=\!0.33$ GeV and $m_s\!=\!0.5$ GeV. Concerning $T_{\rm dec}$ the values $0.155$ and $0.17$ GeV are explored.

Notice that, while the model allows one to take properly into account the momentum boost given to the final hadron by the light quark flowing with the medium, by construction we do not get sizable modifications of the heavy flavour hadrochemistry, like e.g. an enhanced production of $D_s$ mesons or $\Lambda_c$ baryons, which might occur in nature~\cite{ALICE_Ds} and could be accommodated by a coalescence model. Within our framework, once formed a string is hadronized as in the vacuum, through the excitation -- while stretching -- of $q\overline{q}$ pairs (or diquark-antidiquark pairs, leading in this case the the production of a baryon-antibaryon pair) from the vacuum: 
having a strange quark as an endpoint doesn't necessarily imply giving rise to a $D_s$ meson. Furthermore, while in the case of p-p collisions the PYTHIA parton showers keep track of all color connections and the strings to hadronize are usually quite elongated objects, with a lot of kinks describing the radiated gluons and contributing to the soft hadron multiplicity, here the best we can do is to couple the heavy quark $Q$ to an antiquark from the medium, ignoring what happened before: doing better would require developing a full in-medium parton shower, with a consistent description of collisional and radiative processes. Bearing in mind these limitations (the last one affecting also coalescence models), 
this new recipe has in any case two big advantages with respect to the way we modeled so far hadronization in terms of vacuum fragmentation functions: we can first of all provide a realistic estimate of the role of light quarks to explain peculiar features of the $D$ meson spectra at low and moderate $p_T$; secondly, the complete information on all the final state particles arising from the fragmentation of the strings allows us to provide theory predictions for observables like $D\!-\!h$, $e\!-\!h$, $e^+\!-\!e^-$... correlations to be compared to existing data and possibly used as a guidance to future experimental measurements.

Before displaying the results obtained by interfacing the above model to numerical transport calculations it may be useful to analyze how it compares with other mechanisms of in-medium hadronization, like in particular coalescence. The latter is usually pictured as a $2\to 1$ process in which a quark-antiquark pair from the medium gives rise to a meson $M$, which can be either a stable or a resonant state. Microscopic calculations have been proposed to describe the process, either based on the overlap of the parton and meson wave-functions~\cite{rapp} or on the Boltzmann equation with $q\overline{q}\to M$ inelastic collisions~\cite{ravagli}.
In our approach we don't develop a full quantum-mechanical calculation and we rather assume that the hadronizing heavy quark will always find a thermal parton nearby to give rise to a colour-singlet object of arbitrary invariant mass (which does not have to correspond to any particular resonance). We identify such a color-singlet state with a string and let it fragment into the final hadrons according to the Lund model implemented in PYTHIA. Hence, what we consider are $2\to n$ process (e.g. $Q+\overline{q}\to D\pi\pi$). At the end in any case, both with coalescence and with our model, the net effect is that the light thermal parton will transfer his additional radial and elliptic flow to the final charmed hadron. 
\subsection{Transport calculations: $R_{AA}$ and $v_2$}\label{sec:results}
After discussing the limiting scenario of full kinetic thermalization of heavy flavours here we consider the predictions of our POWLANG transport model, which should provide a description closer to reality, where heavy quarks exchange momentum at a finite rate with a medium which expands and cools, losing energy and tending to inherit part of the flow of the latter.
In particular in this section we display the results obtained by applying the string-based hadronization mechanism for heavy quarks previously described to the transport calculations performed with our POWLANG setup.
Our transport model has been presented in detail in previous publications~\cite{lange,lange2} and here we just recall its essential points. The initial $Q\overline{Q}$ production in hard pQCD events is simulated through the POWHEG-BOX package, with initial state nuclear effects encoded into EPS09 modifications of the PDFs~\cite{eps}. 
Heavy quarks are then distributed in the transverse plane according to the local density of binary nucleon-nucleon collisions, which is taken from an optical Glauber calculation, used also to initialize the (2+1)D hydrodynamic calculations, performed with the code by Luzum and Romatschke~\cite{rom1,rom2} describing the medium evolution. 
The heavy quark propagation in the medium is governed by a relativistic Langevin equation, with transport coefficients either obtained through weak-coupling calculations (with Hard Thermal Loop resummation of medium effects) or based on lattice QCD simulations~\cite{lat1,lat2,lat3}. Heavy quark hadronization (either via the usual in-vacuum fragmentation functions or through the new in-medium string-fragmentation routine) is performed once, during its propagation, the heavy quark is found in a cell below a given decoupling temperature $T_{\rm dec}$.

In this section we will focus on a few heavy flavour observables, in particular the nuclear modification factor $R_{AA}$ and the elliptic flow $v_2$, in the regime of low and moderate $p_T$ where the new hadronization scheme introduces important qualitative changes to the results.

\begin{figure}[!h]
\begin{center}
\includegraphics[clip,width=0.48\textwidth]{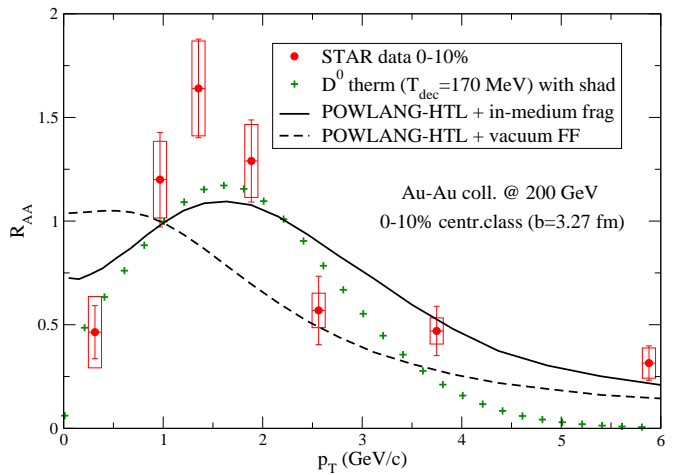}
\caption{$R_{AA}$ of $D^0$ mesons in central Au-Au collisions at $\sqrt{s_{NN}}\!=\!200$ GeV. POWLANG results obtained with HTL transport coefficients and a decoupling temperature $T_{\rm dec}\!=\!170$ MeV are displayed. 
The new in-medium hadronization mechanism leads to a characteristic bump in the $R_{AA}$ due to the radial flow of light quarks. Also shown for comparison is the limiting case of full kinetic thermalization of $D$ mesons. Theory curves are compared to STAR data~\cite{STARD}.}\label{fig:RAA_D_RHIC_transp} 
\end{center}
\end{figure}
\begin{figure}[!h]
\begin{center}
\includegraphics[clip,width=0.48\textwidth]{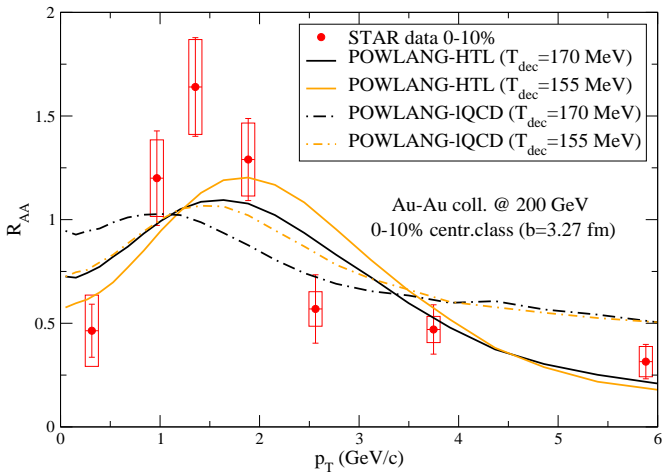}
\caption{$R_{AA}$ of $D^0$ mesons in central Au-Au collisions at $\sqrt{s_{NN}}\!=\!200$ GeV. POWLANG results are displayed for two different choices of transport coefficients (HTL, continuous curves, and l-QCD, dot-dashed curves ) and decoupling temperatures ($T_{\rm dec}\!=\!170$ MeV, in black, and $155$ MeV, in orange). In all cases heavy quarks are hadronized via recombination with light partons forming Lund strings eventually fragmented. Theory curves are compared to STAR data~\cite{STARD}.}\label{fig:RAA_D_RHIC_HTLvsLAT} 
\end{center}
\end{figure}
In Fig.~\ref{fig:RAA_D_RHIC_transp} we start displaying some POWLANG outcomes for the $R_{AA}$ of $D^0$ mesons in central ($0-10\%$) Au-Au collisions at $\sqrt{s_{NN}}\!=\!200$ GeV. HTL transport coefficients are employed and the difference between the two hadronization schemes (here taken to occur at $T=170$ MeV), either with vacuum fragmentation or with in-medium string fragmentation, are clearly visible: in the second case the radial flow of the light thermal parton leads to the development of a bump around $p_T\sim 1.5$ GeV in qualitative agreement with the experimental data. 
Also shown for comparison is the limit of complete kinetic thermalization previously discussed, which displays a behaviour close to the tranport results with in-medium string fragmentation at moderate $p_T$, the bump in the $R_{AA}$ being just a bit sharper: at larger $p_T$ of course the thermal spectrum drops to zero more quickly. 
In Fig.~\ref{fig:RAA_D_RHIC_HTLvsLAT} we address the same data performing a more systematic theoretical analysis. Results obtained with two different sets of tranport coefficients, HTL (continuous curves) and lattice-QCD (dot-dashed curves), and with two different hadronization temperatures for charm, $T_{\rm dec}=170$ MeV (black curves) and $T_{\rm dec}=155$ MeV (orange curves) are compared. Notice how, at large $p_T$, the exact value of the decoupling temperature is not important, while at low $p_T$ it plays a non negligible role, since for lower $T_{\rm dec}$ the system has time to develop a larger radial flow. The value $T_{\rm dec}=155$ MeV is more in line with the most recent lattice-QCD estimates of the pseudo-critical temperature~\cite{latticeTc}, while the choice $T_{\rm dec}=170$ MeV may be justified by the possibility that the quark-to-hadron transition occurs earlier for heavier particles~\cite{latticemass}: we think it is fair to consider this as part of the systematic uncertainty. Notice furthermore that we are neglecting any rescattering of the $D$ mesons in the hadronic phase, which may also affect the final spectra.

\begin{figure}[!h]
\begin{center}
\includegraphics[clip,width=0.48\textwidth]{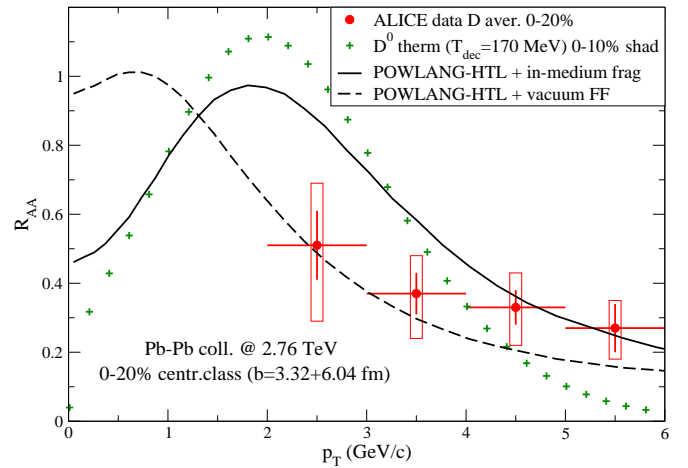}
\caption{The same as in Fig.~\ref{fig:RAA_D_RHIC_transp}, but for central Pb-Pb collisions at $\sqrt{s_{NN}}\!=\!2.76$ TeV. POWLANG results with in-vacuum and in-medium HQ fragmentation and decoupling temperature $T_{\rm dec}=170$ MeV are compared to ALICE data in central ($0-20\%$) collisions~\cite{ALICE_DRAA}. Also shown is the limit of full kinetic thermalization, with Cooper-Frye decoupling at $T_{\rm dec}$.}\label{fig:RAA_D_LHC_transp} 
\end{center}
\end{figure}
\begin{figure}[!h]
\begin{center}
\includegraphics[clip,width=0.48\textwidth]{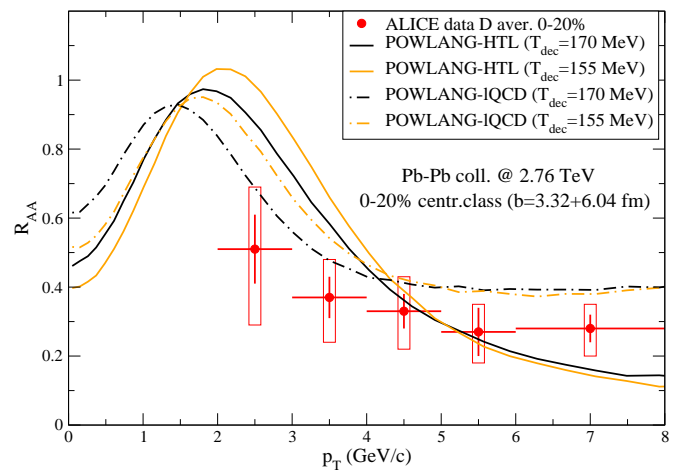}
\caption{The same as in Fig.~\ref{fig:RAA_D_RHIC_HTLvsLAT}, but for central Pb-Pb collisions at $\sqrt{s_{NN}}\!=\!2.76$ TeV.}\label{fig:RAA_D_LHC_HTLvsLAT}
\end{center}
\end{figure}
In Figs.~\ref{fig:RAA_D_LHC_transp} and \ref{fig:RAA_D_LHC_HTLvsLAT} we performed the same study for $D^0$ mesons in central ($0-20\%$) Pb-Pb collisions at the LHC at $\sqrt{s_{NN}}\!=\!2.76$ TeV. In this case the agreement between the outcomes of transport calculations and ALICE experimental data is less satisfactory. In particular present data don't show evidence of a possible bump due to radial flow found by STAR and predicted by our in-medium hadronization mechanism. 
The trend of the data is perhaps best reproduced by the simulations with lattice-QCD transport coefficients, although the data show a larger suppression in the range $p_T\!>\!2$ GeV/c.
More experimental data at low $p_T$ are necessary in order to draw conclusions abut the existence of an enhancement in the $R_{AA}$ arising from radial flow.
Notice that no medium modification of the heavy flavour hadrochemistry has been included in our fragmentation routine; due to the (almost exact, during the limited lifetime of the fireball) conservation of the total number of charm quarks, a possible increased production of $D_s$ mesons or $\Lambda_c$ baryons in A-A collisions would entail a corresponding depletion of the yields of the other charmed hadrons which, limiting the study to non-strange $D$-mesons, could lead to a reduction of the observed $R_{AA}$.

\begin{figure}[!h]
\begin{center}
\includegraphics[clip,width=0.48\textwidth]{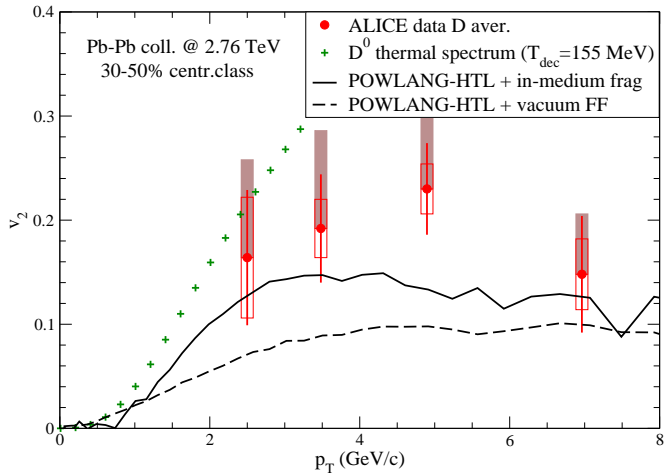}
\caption{$D$-meson $v_2$ in Pb-Pb collisions at $\sqrt{s_{NN}}\!=\!2.76$ TeV. POWLANG results (with HTL transport coefficients) with in-vacuum and in-medium HQ fragmentation at the decoupling temperature $T_{\rm dec}=155$ MeV are compared to ALICE data in the 30-50\% centrality class~\cite{ALICE_Dv2} and to the limit of kinetic thermalization. The new hadronization mechanism, with the fragmentation of $Q\overline{q}_{\rm therm}$ strings, leads to a sizable increase of the elliptic flow of $D$ mesons.}\label{fig:v2_D_LHC_transp} 
\end{center}
\end{figure}

\begin{figure}[!h]
\begin{center}
\includegraphics[clip,width=0.48\textwidth]{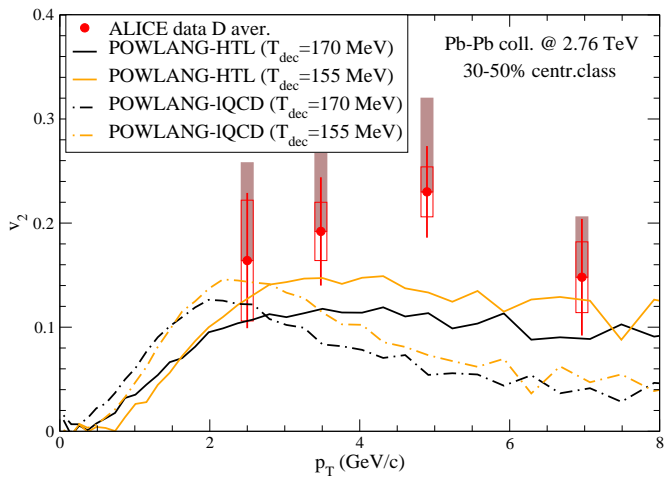}
\caption{$D$ meson $v_2$ in Pb-Pb collisions at $\sqrt{s_{NN}}\!=\!2.76$ TeV. POWLANG results are displayed for two different choices of transport coefficients (HTL, continuous curves, and l-QCD, dot-dashed curves ) and decoupling temperatures ($T_{\rm dec}\!=\!170$ MeV, in black, and $155$ MeV, in orange). In all cases heavy quarks are hadronized via recombination with light partons forming Lund strings eventually fragmented. Theory curves are compared to ALICE data in the $30-50\%$ centrality class~\cite{ALICE_Dv2}.}\label{fig:v2_D_LHC_HTLvsLQCD} 
\end{center}
\end{figure}
In Figs.~\ref{fig:v2_D_LHC_transp} and \ref{fig:v2_D_LHC_HTLvsLQCD} we address the $v_2$ of $D$ mesons in semicentral (30-50\%) Pb-Pb collisions at $\sqrt{s_{NN}}\!=\!2.76$ TeV at the LHC, comparing our model calculations to the results obtained by the ALICE Collaboration~\cite{ALICE_Dv2}: in the figures the shaded boxes represent the systematic uncertainty due to the subtraction of the $B$ feed-down contribution. The elliptic flow measurement has the virtue of being insensitive to the systematic theoretical uncertainty on the absolute normalization of the various hadron spectra (arising from the poorly constrained nPDFs and from the possible change in the hadrochemistry in A-A collisions) and of providing a complementary information on the medium with respect to the $R_{AA}$: it is in fact more sensitive to the latest stages of the fireball evolution, when the latter has acquired a larger elliptic flow (initially vanishing), while most of the energy-loss occurs on the contrary in the earliest stages when the medium is denser.  
In Fig.~\ref{fig:v2_D_LHC_transp} we show the effect of new the procedure for in-medium hadronization through string fragmentation. The figure refers to the case of HTL transport coefficients, with a decoupling temperature $T_{\rm dec}=155$ MeV. While POWLANG outcomes with standard in-vacuum fragmentation of charm largely underpredicts the observed $v_2$, the additional flow acquired from the light thermal partons move the theory curves with in-medium hadronization closer to the experimental data. 
In  Fig.~\ref{fig:v2_D_LHC_HTLvsLQCD} we present a more systematic study of the various scenarios, comparing different choices of transport coefficients (HTL, continuous curves, and lattice-QCD, dot-dashed curves) and decoupling temperatures $T_{\rm dec}\!=\!170$ MeV (black curves) and $T_{\rm dec}\!=\!155$ MeV (orange curves). The best agreement with the current experimental data is obtained with HTL transport coefficients and a hadronization temperature of 155 MeV: both the heavy quarks and the light thermal partons to which they are recombined have more time in this case to develop an elliptic flow. Lattice-QCD coefficients (much larger than weak coupling ones at zero momentum) provide a larger $v_2$ at low $p_T$; however, neglecting their possible momentum dependence leads to underestimate the elliptic flow at high $p_T$

\begin{figure}[!h]
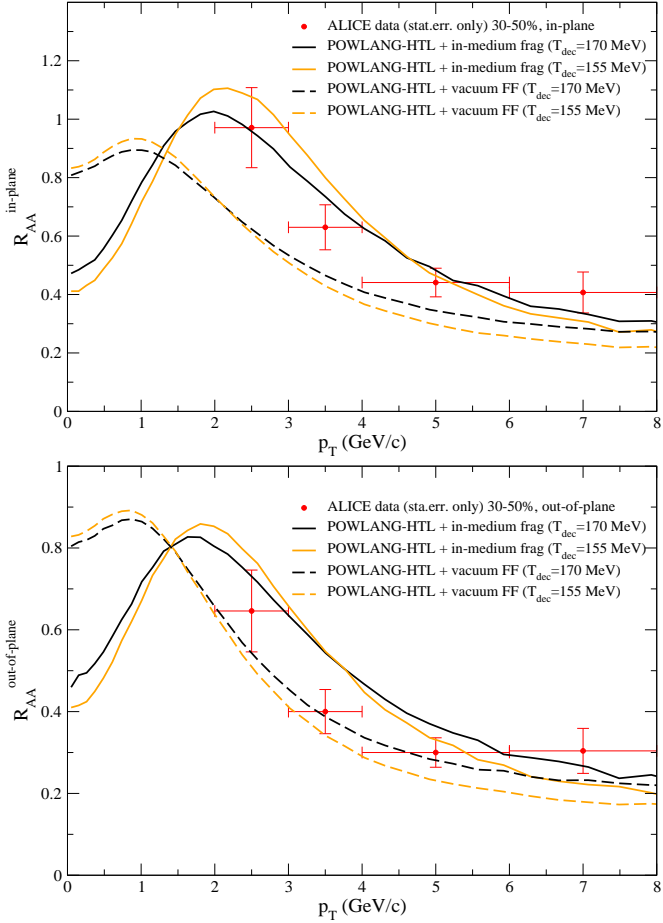

\begin{center}
\includegraphics*[clip,width=0.48\textwidth]{RAA_inplane_medvsvac.eps}
\includegraphics*[clip,width=0.48\textwidth]{RAA_outofplane_medvsvac.eps}
\caption{$R_{AA}$ in-plane (upper panel) and out-of-plane (lower panel) of $D$ mesons. ALICE data in the $30\!-\!50\%$ centrality class~\cite{ALICE_reactionplane} are compared to POWLANG results obtained with different hadronization mechanisms (in-medium, solid curves, and vacuum fragmentation, dashed curves) and decoupling temperatures ($T_{\rm dec}\!=\!170$ MeV, black curves, and $155$ MeV, orange curves).}\label{fig:RAA_inout_medvac} 
\end{center}
\end{figure}

\begin{figure}[!h]
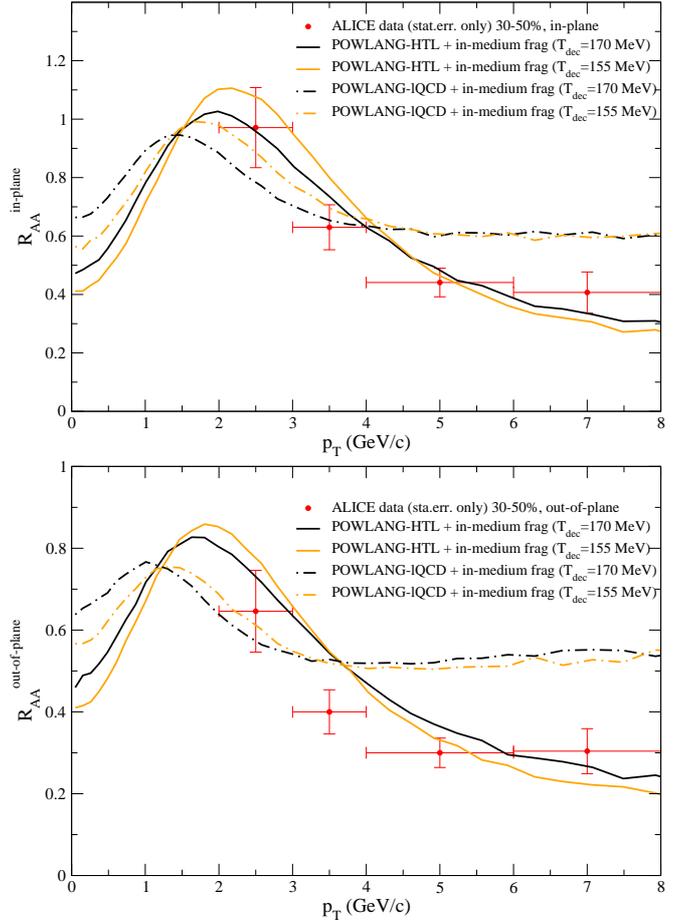

\begin{center}
\includegraphics*[clip,width=0.48\textwidth]{RAA_inplane_HTLvsLat.eps}
\includegraphics*[clip,width=0.48\textwidth]{RAA_outofplane_HTLvsLat.eps}
\caption{$R_{AA}$ in-plane (upper panel) and out-of-plane (lower panel) of $D$ mesons. ALICE data in the $30\!-\!50\%$ centrality class~\cite{ALICE_reactionplane} are compared to POWLANG results obtained with different transport coefficients (HTL, continuous curves, and lQCD, dashed curves) and decoupling temperatures ($T_{\rm dec}\!=\!170$ MeV, black curves, and $155$ MeV, orange curves).}\label{fig:RAA_inout_HTLlQCD} 
\end{center}
\end{figure}

For the same $30\!-\!50\%$ centrality class the ALICE Collaboration~\cite{ALICE_reactionplane} has recently provided also the results for the $D$-meson $R_{AA}$ in-plane ($-\pi/4\!\le\!\phi\!\le\!\pi/4$ and $3\pi/4\!\le\!\phi\!\le\! 5\pi/4$ with respect to the reaction plane) and out-of-plane (in the other two quarters). This represents a way of combining the information on the quenching of the spectra (usually studied via the azimuthally integrated nuclear modification factor) and on the azimuthal anisotropy of particle production (usually studied via the elliptic-flow coefficient $v_2$, without the need of any p-p reference).
In Figs.~\ref{fig:RAA_inout_medvac} and~\ref{fig:RAA_inout_HTLlQCD} we display the outcomes of POWLANG simulations for various choices of the hadronization mechanism (in-medium or in-vacuum fragmentation), of the tranport coefficients (weak coupling \mbox{HTL} or lattice-QCD) and of the decoupling temperature. While out-of-plane data are not able to discriminate between in-medium and in-vacuum hadronization, the in-plane $R_{AA}$ is better reproduced by our new routine based on the formation of strings with light thermal partons. Concerning the transport coefficients, the lattice-QCD results display a too small quenching at high $p_T$, due to the absence of information on their momentum dependence. Data at low $p_T$ would be clearly needed to exctract more solid information both on heavy flavour hadronization and transport coefficients: at the current stage, at variance with more central collisions, experimental results seem quite well reproduced employing HTL transport coefficients and forcing heavy quarks to form strings with light partons from the medium at hadronization.


\subsection{Heavy-flavour azimuthal correlations}\label{sec:corr}
\begin{figure}[!h]
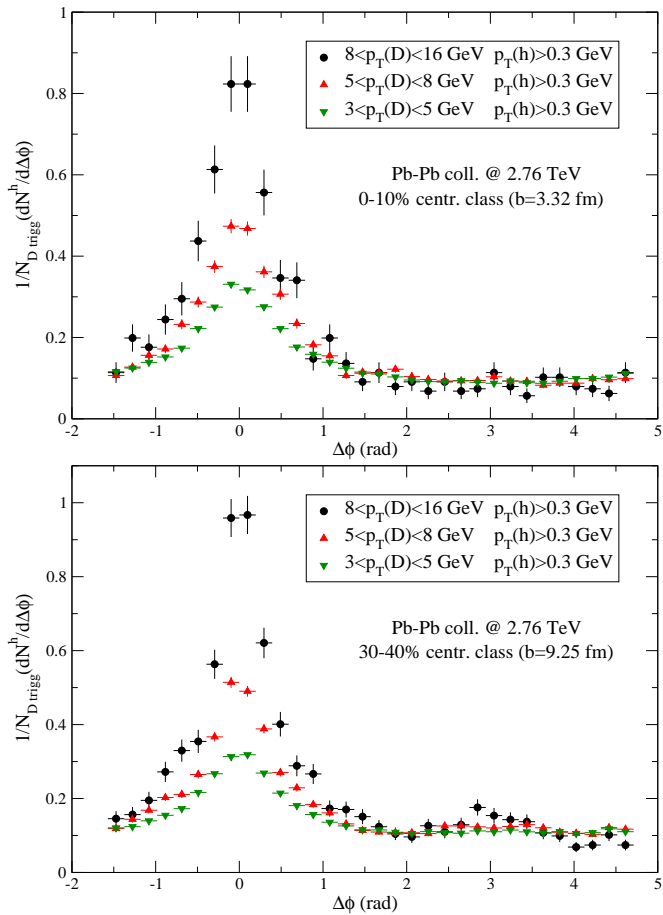

\begin{center}
\includegraphics*[clip,width=0.48\textwidth]{D-h_corr_PbPb_0-10_Tdec170.eps}
\includegraphics*[clip,width=0.48\textwidth]{D-h_corr_PbPb_30-40_Tdec170.eps}
\caption{$D\!-\!h$ correlations in Pb-Pb collisions at $\sqrt{s_{\rm NN}}\!=\!2.76$ TeV for various $p_T$ cuts on the trigger particle and two different centrality classes: $0-10\%$ (upper panel) and $30-40\%$ (lower panel).}\label{Dhcorr} 
\end{center}
\end{figure}
\begin{figure}[!h]
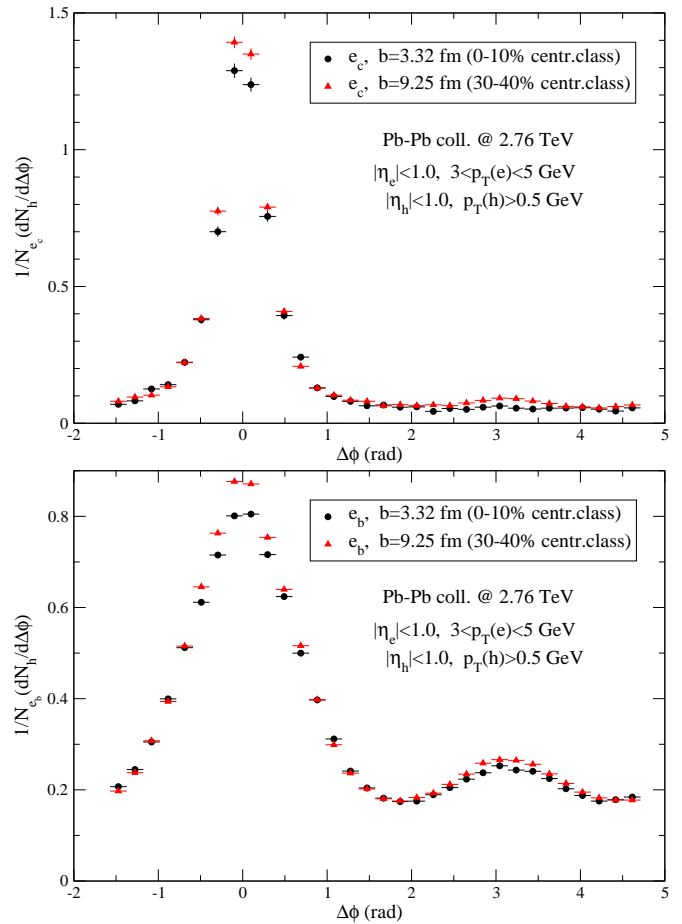

\begin{center}
\includegraphics*[clip,width=0.48\textwidth]{ec-h_corr_vs_centr_pT3-5_softass_elresampl.eps}
\includegraphics*[clip,width=0.48\textwidth]{eb-h_corr_vs_centr_pT3-5_softass.eps}
\caption{$e\!-\!h$ correlations in Pb-Pb collisions at $\sqrt{s_{\rm NN}}\!=\!2.76$ TeV for various centrality classes: $0-10\%$ (black circles) and $30-40\%$ (red triangles). The separate charm (upper panel) and beauty (lower panel) results are shown.}\label{ehcorr} 
\end{center}
\end{figure}
\begin{figure}[!h]
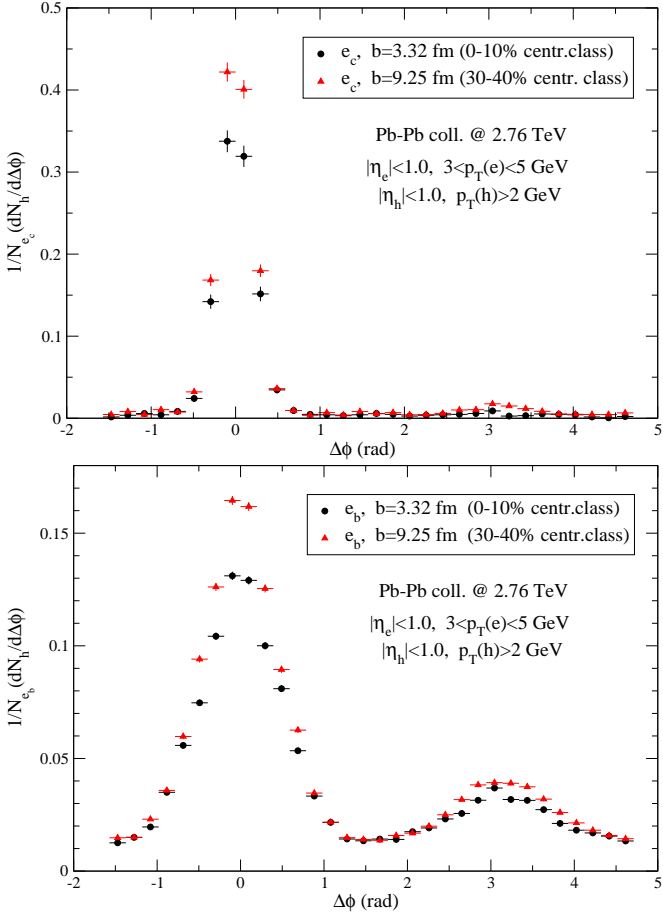

\begin{center}
\includegraphics*[clip,width=0.48\textwidth]{ec-h_corr_vs_centr_pT3-5elresampl.eps}
\includegraphics*[clip,width=0.48\textwidth]{eb-h_corr_vs_centr_pT3-5.eps}
\caption{The same as in Fig.~\ref{ehcorr} for harder $p_T$ cuts on the associated hadrons.}\label{ehcorr_harder} 
\end{center}
\end{figure}
The full differential information on the final state provided by our setup -- from the simulation of the initial hard production of the $Q\overline{Q}$ pairs, to the modeling of their propagation in the plasma and of their hadronization -- opens the possibility of providing predictions for more differential observables like angular correlations with respect to heavy-flavour trigger particles. The interaction with the medium is expected to alter the momentum and the relative angle of heavy quarks produced in the same hard event and this should be observed in their fragmentation and decay products. Since, due to the small branching ratio into the channel used for their reconstruction, direct $D\!-\!\overline{D}$ correlations are outside the conceivable experimental capabilities, here we focus on more indirect observables such as $D\!-\!h$ and $e\!-\!h$ correlations: the corresponding results are shown in Figs.~\ref{Dhcorr}-\ref{ehcorr_harder}. All figures are obtained with weak-coupling HTL transport coefficients in the QGP phase. Although some theoretical work was already done in the past attempting to quantify medium effects on $Q\overline{Q}$ correlations~\cite{nantescorr}, here we push theory predictions until the actual experimental observables accessible in the near future, namely angular correlations between heavy flavour particles (or their decay products) and the charged hadrons produced in the same collision. In general one observes a strong suppression of the away-side peak around $\Delta\phi\!=\!\pi$. Depending on the cuts imposed on the trigger particles (either $D$-mesons or heavy-flavour decay electrons) and on the associated hadrons this can be mostly due either to the energy loss (moving particles below the $p_T$-cut) or to the angular decorrelation (moving particles away from $\Delta\phi\!=\!\pi$) of the parent heavy quark. Concerning the $e\!-\!h$ correlations we have plotted separately the charm and beauty contributions. While in the case of electrons from charm decays we have always found an almost complete suppression of the away-side peak independently on the centrality of the collision and on the kinematic cuts, beauty decay electrons turn out to be less decorrelated, allowing one in principle to extract more information on the heavy-quark interaction with the medium.  

\begin{figure}[!h]
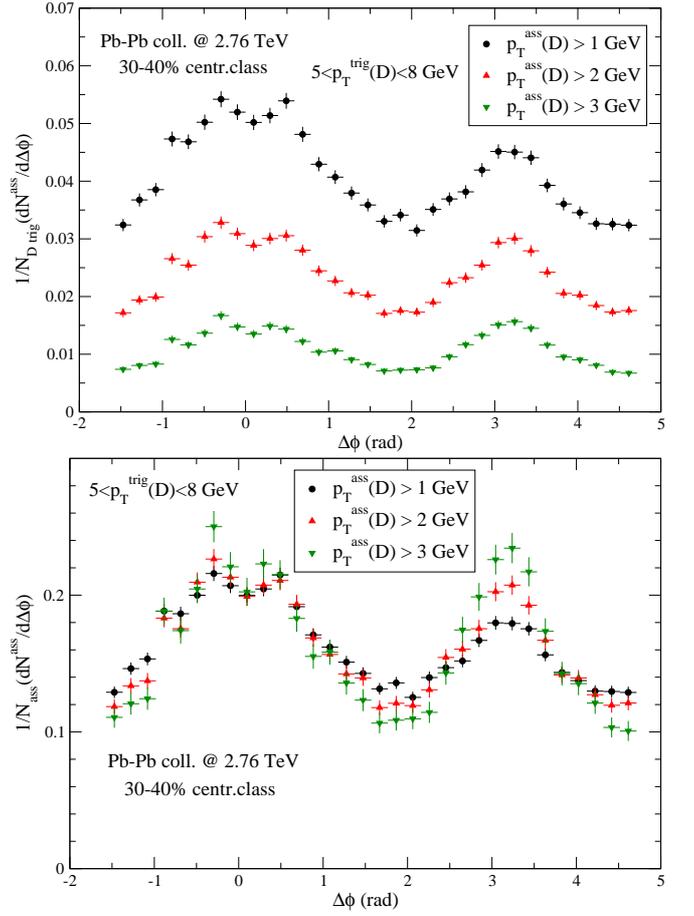

\begin{center}
\includegraphics[clip,width=0.48\textwidth]{D-Dbar_corr_PbPb_30-40_Tdec170.eps}
\includegraphics[clip,width=0.48\textwidth]{D-Dbar_corr_PbPb_30-40_Tdec170_norm.eps}
\caption{$D\!-\!\overline{D}$ azimuthal correlations in semi-central Pb-Pb collisions (30-40\%) at $\sqrt{s_{\rm NN}}\!=\!2.76$ TeV for different $p_T$-cuts on the associated particle. In the lower panel all the distributions are normalized to unit area, in order to enlight only the effect of the broadening of the aways-side peak.}\label{fig:DDbar} 
\end{center}
\end{figure}
Finally, in spite of being presently -- as above mentioned -- something beyond the experimental capability, we consider useful to check what would be the results for the $D\!-\!\overline{D}$ correlations provided by our model: an example is shown in Fig.~\ref{fig:DDbar}. The most evident feature, at variance with the $D\!-\!h$ and $e\!-\!h$ cases, is the presence in our results of two well defined peaks, not washed-out by the interaction of the heavy quarks with the plasma. The near-side peak arises from the $Q\overline{Q}$ pairs produced close in angle via gluon-splitting. In the upper panel of Fig.~\ref{fig:DDbar} we display model predictions for $D\!-\!\overline{D}$ correlations in semi-central Pb-Pb collisions (30-40\%) at $\sqrt{s_{\rm NN}}\!=\!2.76$ TeV for different $p_T$-cuts on the associated particle in the pseudorapidity window $|\eta_D|<1$.
For all the considered cases the trigger $D$ meson is taken in the same $p_T$ range $5\!<\!p_T^{\rm trig}\!<\!8$ GeV/c. The softer associated particles typically is the one coming from the quark of the pair which has crossed the longer medium thickness, losing energy and received random kicks from the plasma which make it deviate from its initial trajectory (leading to a broadening of the away-side peak). Due to energy loss, for harder requirements on the $p_T$ of the associated $D(\overline{D})$ meson the distribution gets more and more quenched, with less and less charmed hadrons (their yields being given by the integral of the curves) satisfying the kinematic cuts.  
Besides the above medium effects there are of course other possible sources of momentum unbalance and angular decorrelation arising both from the fragmentation of the quarks into the final $D$ mesons and from the hard event itself, in which (typically softer) gluons can be produced together with the $Q\overline{Q}$ pairs: of course these effects are also present in p-p collisions.   

In order to isolate the genuine angular decorrelation effect from the one due to energy loss, in the lower panel we normalize all the distributions to unit area so that one can appreciate how the away-side peak gets broader for softer $p_T$-cuts. The fact that for harder $p_T$ selection on the associated particle the peak remains sharper might reflect the flat momentum dependence of the $\kappa_T$ transport coefficients predicted by weak coupling calculations and employed in the simulations. Notice that the pedestal is given by $D-\overline{D}$ pairs which have been completely decorrelated in angle.

Why do model predictions for $D\!-\!h$ and $e\!-\!h$ correlations display on the contrary such a dramatic unbalance between the two peaks? Comparing for instance Figs.~\ref{Dhcorr} and~\ref{fig:DDbar} one observes in the first case, for the same centrality class and kinematic cuts on the trigger particle, a much more pronounced near-side peak. The latter receives several different contributions: from the fragments of the string of the leading $Q$; from the fragments of the string of an associated $\overline{Q}$ close in angle because arising from gluon splitting and also, in this case, from its charged decay hadrons which are not subtracted. Concerning the $D-\overline{D}$ correlations, on the contrary, the only contribution to the near-side peak is from $c\overline{c}$ pairs from gluon splitting. The area below the near-side peak is in this case much smaller compared to Fig.~\ref{Dhcorr} and in particular is now of the same order as the one on the away-side. Of course each associated $\overline{D}$ will decay into two or three hadrons, but part of them will be too soft to satisfy the kinematic requirements imposed in Fig.~\ref{Dhcorr} and the ones within the cuts will not be sufficient to provide a significant contribution to the near-side peak: the origin of the latter must be then attributed to hadrons from string fragmentation, with the most relevant contribution provided presumably by the one attached to the trigger heavy quark, always present and boosted to large momentum.  
Concerning the away-side peak, its quenching in Figs.~\ref{Dhcorr}-~\ref{ehcorr_harder} reflects in part the further decorrelation introduced by the decays into light hadrons and in part is a graphical effect arising from the comparison with the enhanced near-side one. In fact looking carefully at the results in the 30-40\% centrality class one can observe some tracks of azimuthal correlation surviving and being of the same order of magnitude of the $D\!-\!\overline{D}$ case. We postpone a more detailed analysis to a forthcoming publication.     
\section{Discussion and outlook}\label{sec:discussion}
In this paper we have developed a simple model to describe heavy quark hadronization in the presence of a hot deconfined medium like the one expected to be formed in high energy nuclear collisions (a Quark-Gluon Plasma). The model is based on the fragmentation of strings formed combining a heavy quark with a light thermal parton from the plasma. Such a model has been implemented into a numerical routine interfaced with our previously developed POWLANG transport code allowing the simulation of the production and of the propagation in the plasma of $Q\overline{Q}$ pairs in  high-energy nucler collisions. Results have been shown for the $R_{AA}$ (considering also the in-plane and out-of-plane cases) of $D$-mesons in Au-Au and Pb-Pb collisions at RHIC and LHC respectively: in general the improved model provides a better description of the experimental data.

In parallel, thanks also to the new model of hadronization at our disposal, we have tried to answer the question of the possible kinetic equilibration of charm in heavy-ion collisions. We have shown how predictions of two different scenarios ({hydrodynamic limit} and {transport + recombination}) {can be hardly distinguished} within the kinematic range covered by the current experimental results and they can {both describe} some {qualitative features of the data} at low and moderate $p_T$.

In the final part of the paper we have shown the potentiality of our setup for the study of heavy-flavour correlations in heavy-ion collisions. We consider this just an exploratory study, a deeper theoretical analysis being necessary in order to understand which additional information on the heavy quark interaction with the medium can be obtained in principle from such experimental measurements: we leave such an issue for future work.



\begin{thebibliography}{99}
\bibitem{PHEe} PHENIX Collaboration (A. Adare \emph{et al.}),
Phys. Rev. Lett. 98, 172301 (2007) and Phys. Rev. C 84, 044905 (2011).
\bibitem{STARe} STAR Collaboration (M. Mustafa for the collaboration),
Nucl. Phys. A 904-905 (2013) 665c.
\bibitem{ALICEe} ALICE Collaboration (S. Sakai for the collaboration),
Nucl. Phys. A 904-905 (2013) 661c.
\bibitem{ALICEmu} ALICE Collaboration (D. Stocco for the collaboration),
Nucl. Phys. A 910-911 (2013) 355. 
\bibitem{ALICE_DRAA} ALICE Collaboration (B. Abelev \emph{et al}.),
JHEP 1209 (2012) 112.
\bibitem{ALICE_Dv2} ALICE Collaboration (B. Abelev \emph{et al}.),
Phys. Rev. Lett. 111 (2013) 102301.
\bibitem{CMS_Jpsi} CMS Collaboration (S. Chatrchyan \emph{et al}.),
JHEP 1205 (2012) 063 and CMS-PAS-HIN-12-014. 
\bibitem{STARD} STAR Collaboration (L. Adamczyk \emph{et al.}),
arXiv:1404.6185 [nucl-ex].
\bibitem{aic} P.B. Gossiaux and J. Aichelin,
Phys. Rev. C 78 (2008) 014904.
\bibitem{aic2} P.B. Gossiaux, R. Bierkandt, J. Aichelin,
Phys. Rev. C {79}, 044906 (2009).
\bibitem{BAMPS}  J. Uphoff \emph{et al.},
Phys. Rev. C 84 (2011) 024908. 
\bibitem{BAMPS2} 
J. Uphoff, O. Fochler, Z, Xu and C. Greiner,
Phys. Lett. B 717 (2012) 430.
\bibitem{tea} G.D. Moore, D. Teaney,
Phys. Rev. C 71, 064904 (2005).
\bibitem{aka} Y. Akamatsu, T. Hatsuda, T. Hirano,
Phys. Rev. C 79, 054907 (2009).
\bibitem{rapp} H. van Hees {et al.},
Phys. Rev. Lett. 100 (2008) 192301. 
\bibitem{rapp2} 
M. He, R.J. Fries and R. Rapp,
Phys.Rev. C86 (2012) 014903.
\bibitem{lange0} A. Beraudo, A. De Pace, W.M. Alberico and A. Molinari,
Nucl. Phys. A 831 (2009) 59.
\bibitem{lange} W.M. Alberico \emph{et al.}, 
Eur. Phys. J. C 71 (2011) 1666.
\bibitem{lange2} W.M. Alberico \emph{et al.}, 
Eur. Phys. J. C 73 (2013) 2481. 
\bibitem{POW} S. Frixione, P. Nason, G. Ridolfi,
                 JHEP {0709} (2007) 126.
\bibitem{POWBOX} S. Alioli, P. Nason, C. Oleari and E. Re,
JHEP {1006} (2010) 043.
\bibitem{PYTHIA} T. Sjostrand, S. Mrenna and P.Z. Skands,
JHEP 0605 (2006) 026.
\bibitem{sandro} S. Bjelogrli\'c (on behalf of the ALICE Collaboration), Quark Matter 2014 proceedings.
\bibitem{rossi} ALICE Collaboration (A. Rossi for the collaboration), arXiv:1409.4001 [hep-ex] and private communications.
\bibitem{deepa} ALICE Collaboration (D. Thomas for the collaboration),
J.Phys.Conf.Ser. 509 (2014) 012079, arXiv:1312.1489 [hep-ex].
\bibitem{rom1} P. Romatschke, U. Romatschke,
               Phys. Rev. Lett. 99, 172301 (2007) 

\bibitem{rom2} M. Luzum, P. Romatschke,
               Phys. Rev. C {78}, 034915 (2008).
\bibitem{lat1} D. Banerjee \emph{et al.}, Phys.Rev.  D85 (2012) 014510.
\bibitem{lat2} A. Francis \emph{et al.}, PoS LATTICE2011 202.
\bibitem{lat3} A. Francis \emph{et al.} arXiv:1311.3759 [hep-lat].
\bibitem{ber} A. Beraudo, arXiv:1407.5918 [hep-ph].
\bibitem{cooper} F. Cooper and G. Frye,
Phys. Rev. D 10 (1974) 186.
\bibitem{zaida} ALICE Collaboration (Zaida Conesa del Valle for the collaboration),
Nucl. Phys. A 904-905 (2013) 178c.
\bibitem{ALICE_Ds} ALICE Collaboration (G.M. Innocenti for the collaboration),
Nucl. Phys. A 904-905 (2013) 433c.
\bibitem{ravagli} L. Ravagli, H. van Hees and R. Rapp
Phys. Rev. C 79 (2009) 064902. 
\bibitem{eps} K.J. Eskola, H. Paukkunen, C.A. Salgado,
              JHEP 0904 (2009) 065.
\bibitem{latticeTc} Wuppertal-Budapest Collaboration JHEP 1009 (2010) 073
\bibitem{latticemass} R. Bellwied \emph{et al.}, Phys.Rev.Lett. 111 (2013) 202302.
\bibitem{ALICE_reactionplane} ALICE Collaboration (B. Abelev \emph{et al}.),
arXiv:1405.2001 [nucl-ex].
\bibitem{nantescorr} 
M. Nahrgang, J. Aichelin, P.B. Gossiaux and K. Werner,
Phys. Rev. C 90 (2014) 024907.
\end{thebibliography}
\end{document}